\documentclass[preprint2]{aastex63}
\newcommand{\kms}{{km s$^{-1}$}}

\newcommand{\myemail}{sao@astro.snu.ac.kr}
\newcommand{\profemail}{mglee@astro.snu.ac.kr}

\usepackage{amsmath}
\usepackage{multirow}
\hypersetup{linkcolor=red,citecolor=blue,filecolor=cyan,urlcolor=magenta}

\usepackage{amssymb}

\shorttitle{Globular Clusters in NGC\,4839}
\shortauthors{Oh et al.}

\begin{document}

\title{ %Third Revised Draft 1 \today \\
Globular Clusters in NGC\,4839 Falling into Coma:
Evidence for the Second Infall?}

\correspondingauthor{Myung Gyoon Lee}
\email{\myemail,\profemail}
\author[0000-0003-2288-9093]{Seong-A Oh}
\author[0000-0003-2713-6744]{Myung Gyoon Lee}
\affil{Astronomy Program, Department of Physics and Astronomy, SNUARC, Seoul National University, 1 Gwanak-ro, Gwanak-gu, Seoul 08826, Republic of Korea}

\author[0000-0002-2502-0070]{In Sung Jang}
\affiliation{Department of Astronomy \& Astrophysics, University of Chicago, 5640 South Ellis Avenue, Chicago, IL 60637, USA}

\begin{abstract}
NGC\,4839 is the brightest galaxy (cD) of the NGC\,4839 group at $R\approx 1$ Mpc in the south-west of the Coma cluster, which is known to be falling into Coma.
However, it has been controversial whether it is in the first phase of infall or  in the second phase of infall after passing the Coma center. 
We present a wide field study of globular clusters (GCs) in NGC\,4839 and its environment based on Hyper Suprime-Cam $gr$ images in the Subaru archive.
We compare the GC system of NGC\,4839 with that of NGC\,4816, which is the brightest member (S0) of the nearby group and  lies at a similar distance in the west from the Coma center.  
Interestingly the spatial distribution of the GCs in NGC\,4839 is significantly more compact than that of the GCs in NGC\,4816. In addition, the radial number density profile of the GCs in NGC\,4839 shows an abrupt drop at $R_{N4839}\approx 80$~kpc, while that of the GCs in NGC\,4816 shows a continuous slow decline even in the outer region at $80<R_{N4816}<500$ kpc.
The effective radius of the NGC\,4839 GC system
is about three times smaller than that of the NGC\,4816 GC system.
This striking difference can be explained if NGC\,4839 lost a significant fraction of the GCs in its outskirt when it passed through Coma. This supports strongly the second infall scenario where the NGC\,4839 passed the Coma center about 1.6 Gyr ago, and began the second infall after reaching the apocenter in the south-west recently.  
\end{abstract}

%\keywords{Early-type galaxies (429), Giant Elliptical galaxies (651), Globular star clusters (656), Rich galaxy clusters (2005)}, Galaxy Clusters (584) Intracluster Medium (858} Coma Cluster ()
%Lenticular galaxies (915)} %

\section{Introduction}

\subsection{The NGC\,4839 Group and the Main Cluster in Coma}
Coma is the most massive galaxy cluster in the local universe. It is connected with filaments from neighboring galaxy clusters and hosts various substructures indicating that it is a complex merger system  \citep{col96,mal20,hea21}. Thus, Coma is one of the best targets to study how large scale substructures are assembled and evolve, and has been a focus of many cluster studies in various aspects (see \citet{biv98,chu21} and references therein).
Two most prominent substructures in Coma  are the main cluster core in the center and the NGC\,4839 group in the south-west, as shown by galaxy number density maps \citep{col96,hea21}, X-ray images of hot gas \citep{whi93,neu01,lys19,chu21}, and radio images of synchrotron emission \citep{bon21,bon22,lal22}.
The main cluster core hosts two giant galaxies (NGC\,4874 (cD) and NGC\,4889 (D)), which are merging now.
The NGC\,4839 group
is at %$R\approx 1.3$ 
$R\approx 1$~Mpc in the south-west of Coma, and it is much smaller and less massive than the main cluster core \citep{col96,lys19}. 
%It is considered that 
The NGC\,4839 group is considered to be falling into Coma and that the two systems will merge to form a more massive system in the future (\citet{biv98} and references therein).

\subsection{Merger Scenarios for the NGC\,4839 Group: A Pre-merger or a Post-merger?}

It is generally accepted that the NGC\,4839 group is merging with the main cluster. However, whether it is a pre-merger where  the NGC 4839 group is in the first phase of infall  \citep{bri92,whi93,col96,neu01,aka13} or a post-merger %where the NGC 4839 group is in the second infall after passing the Coma center
\citep{bur94,lys19,chu21} has been controversial \citep{san20,hea21}.

We summarize the observational features related with the merging of the NGC\,4839 group in the previous studies in Table~\ref{table:mfeatures}. These features include several substructures seen in X-ray and radio images, 
an excess of E+A galaxies in the SW region of the cluster,  and substructures found in the spatial distribution and kinematics of galaxies. Each feature can be explained with either the pre-merger scenario or the post-merger scenario.
Recently the post-merger scenario, which can better explain the existence of X-ray/radio substructures (in particular, bridges and streams), appears to be more supported \citep{lys19,chu21,chu22,bon21}.
However, even in the recent discussions of both scenarios based on various observations, \citet{hea21} state that {\it Nevertheless, the question whether the NGC\,4839 group is on its first infall or has already passed through the cluster, remains open}. 

\subsection{Globular Clusters as a Probe}
The halos of massive galaxies in galaxy clusters grow via numerous mergers of less massive galaxies and host a large number of globular clusters (GCs). %which are mostly metal-poor. 
Thus, GCs are an excellent probe for investigating the structure of the outer halos in massive galaxies in the local universe, and they provide % {\color{red} (provide)}
a critical clue for revealing the assembly history of galaxy halos.

In this study, we present a wide field survey of  GCs covering the NGC\,4839 group and its environment, based on the archival Subaru/Hypersuprime-Cam (HSC) $gr$ images. 
The primary goals of this study are to derive wide field number density maps of
GCs  and to use them to constrain the merger scenarios of the NGC\,4839 group. 
We adopt the distance to Coma as 100~Mpc \citep{deg20}.

\subsection{Previous Studies of NGC\,4839 GCs}
 The main host of the NGC\,4839 group is NGC\,4839 ($M_V=-23.1$ mag, $v_h=7338$ \kms), which is an elongated cD galaxy \citep{sch88,ali14}.
There are only two previous studies of the GCs in NGC\,4839.
\citet{mar02} applied the surface brightness fluctuation (SBF) method to estimate indirectly the total number of GCs in several bright Coma galaxies from $r$-band images obtained at the 2.5m Issac Newton Telescope.  They found that NGC\,4839 is the second-most GC-rich in Coma,  
%with $N(GC)=10000\pm2500$ ($R<5'?$ and specific frequency $S_N =7.0\pm1.9$) (similar to NGC\,4889 $N(GC)=9600\pm2600$), 
following NGC\,4874 in their sample. 
Later \citet{jor04} presented a $F450W$($B$)/$F814W$($I$) photometry of GCs in NGC\,4839  based on $HST$/WFPC2 images, deriving the total number of GCs to be $N_{tot}(GC)=3060\pm850$, which is three times smaller than the value given by \citet{mar02}. 
%This value leads to $S_N$(GC) of $5.6\pm1.9$.
%These values will be checked with our study!
These previous studies either covered only the small field of NGC\,4839 or used only one band, so little is known about the GCs in the outskirt of NGC\,4839.
Other previous $HST$ surveys of GCs in Coma covered mainly the main cluster core, and did not cover the NGC\,4839 region \citep{pen11,mad18}.

\section{Data}

Utilizing the Subaru/Hyper Suprime-Cam (HSC) archival $gr$ images from the Subaru Mitaka Okayama Kiso Archive system (SMOKA) \citep{aih19}, \citet{oh23} provided a wide field survey of GCs in the entire Coma cluster. 
At the distance of Coma (100~Mpc), one arcsec (arcmin) corresponds to a linear scale of 484.8~pc (29.1~kpc).
Thus GCs at the distance of Coma %(100 Mpc) 
appear as  point sources in the HSC images.
\citet{oh23} obtained 
photometry of the point sources in the seven HSC fields covering the entire Coma cluster, using DAOPHOT \citep{ste87}.
We adopt the AB magnitudes in the SDSS system.
The limiting magnitude with 50\% completeness of detection derived from artificial star experiments is $r\approx 27.1$~mag.
Detailed description of the detection and photometry of the point sources is given in \citet{oh23},
%{\color{red} Detailed description of the detection and photometry of the point sources is given in \citet{oh22}, 
of which we used the data for NGC\,4839 and its environment in this study. 
We apply the foreground extinction correction using the extinction maps for Coma given in \citet{sch98,sch11}.

\section{Results}

\subsection{NGC\,4839 in Comparison with NGC\,4816}

%(NGC\,4816: $V_{T,0}=12.84\pm0.20, A_V=0.024, M_V=12.84-0.02-35.0=-22.18, B_{T,0}=13.88, (B_{T,0}-V_{T,0})=1.04$, S0, RC3)
%Group S14: 17 members (10 Es, 7 S0s, 2 Is), $<v>=6898$ \kms, and $\sigma_v = 521$ \kms. 

%(NGC\,4839: $V_{T,0}=11.95, A_V=0.024, M_V=11.95-0.024-35.0=-23.07, B_{T,0}=12.83, (B_{T,0}-V_{T,0})=0.88$ cD, SA0, RC3)
%Group S11: 24 members (12 Es, 6 S0s, 2 Is, and 2 Us), $<v>=7621$ \kms, and $\sigma_v = 462$ \kms. 

\begin{figure*}%[hbt!]
%    \centering
%    \includegraphics[scale=0.25]
%    \plottwo{Figs1/NGC4839_GC_sdm.pdf}
    \includegraphics[scale=0.9]{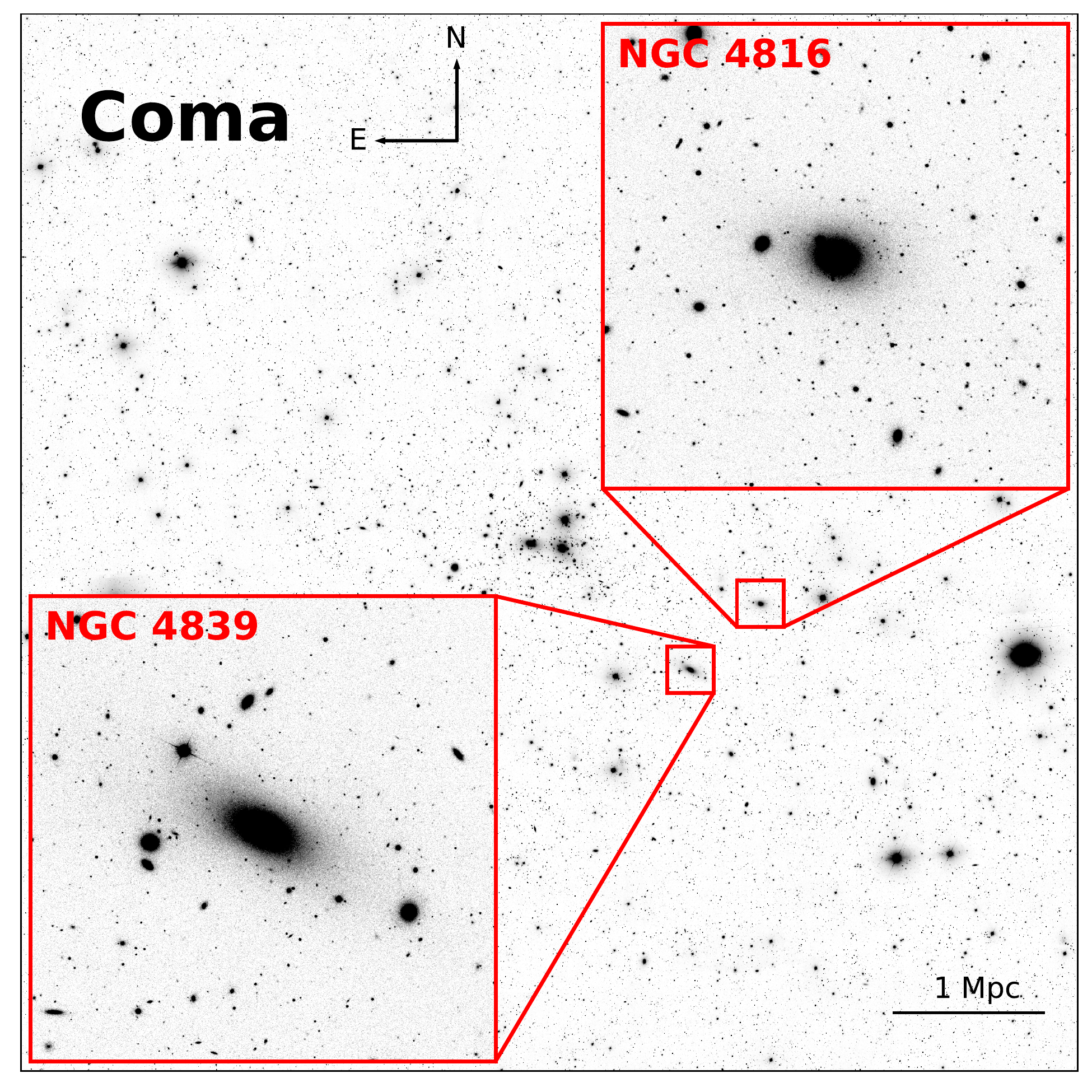} %{fig2a.pdf}   
    %{Figs1/N4839_4816_findmap.pdf} 
    %\includegraphics[scale=0.35]{fig2b.pdf} %Figs1/GC_field167_gal2_figb.pdf}
	\caption{A gray scale map ($4\arcdeg \times 4\arcdeg$) of the $r$-band SDSS image of NGC\,4839 and its environment in the Coma cluster. Zoom-in fields for NGC 4839 and NGC 4816 (red boxes) are $10\arcmin \times 10\arcmin$.
	}
	\label{fig:finder1}
\end{figure*}

In Figure \ref{fig:finder1} %(a)
we show a gray scale map of the $r$-band SDSS  image of the Coma cluster region including the NGC\,4839 group. 
%similar region for comparison. 
The zoom-in  %$r$-band 
images ($10\arcmin \times 10\arcmin$) of NGC\,4839 and NGC\,4816 show that the two galaxies are similar in their luminosity and size.  
In the following analysis of NGC\,4839 we chose NGC\,4816,  a nearby bright S0 galaxy, as a comparison galaxy. % ($M_V=-23.07$ mag, S0, $v_h=7621$ \kms). 

NGC\,4839 and NGC\,4816 are at similar projected distances from the Coma center. The projected separation between the two galaxies in the sky is $21\farcm8$ (0.63~Mpc at the distance of Coma). 
%\textcolor{red}{
\citet{hea21} found 15 groups using the catalog of Coma member galaxies, and provided the number of members and velocity dispersion of each group. Groups S11 and S14 in their study correspond to the NGC 4816 and NGC 4839 groups, respectively. We used these group data in the following analysis. %}

NGC\,4816 is the brightest member  
of the S14 group (with $N(member)=17$ and $\sigma_v = 521$ \kms)
%\footnote{$v =6898$ \kms, and $\sigma_v = 521$ \kms, N(member)=17} 
at $R=49\arcmin$ in the west of Coma (see Fig. 12 in \citet{hea21}).
Similarly, NGC\,4839 is the brightest cD/SA0 member 
of the NGC\,4839 group (the S11 group with $N(member)=24$ and $\sigma_v = 462$ \kms) 
%\footnote{$v=7621$ \kms, and $\sigma_v = 462$ \kms, N(member)=24}) 
at $R=43\arcmin$, but in the south-west of Coma.

%NGC\,4839 is at $R=43\arcmin$ in the south-west and NGC\,4816 at  $R=49\arcmin$ in the west so that 
Thus both galaxies are very bright, and the $V$ magnitude of NGC\,4816 is only 0.9 mag fainter than that of NGC\,4839.
While the NGC\,4839 group shows a strong X-ray emission, the NGC\,4816 group shows little detected X-ray emission even in the recent X-ray images \citep{lys19,san20,mir20,chu21, chu22}.
%NGC\,4839 and NGC\,4816 belong to the same group, G2, based on weak lensing analysis in \citet{oka14}, while 
%NGC\,4839 is included in Group S11 ($v=7621$ \kms, N(member)=24) and NGC\,4816 in Group S14 ($v=6998$ \kms, N(member)=17) based on clustering analysis in \citet{hea21}.

%\textcolor{red}{
Table \ref{table2} lists the basic parameters of the NGC\,4839 and NGC\,4816 groups in comparison with the main cluster.
We calculated the virial mass from the velocity dispersion of the two groups \citep{hea21} 
using the group virial mass equation: $M_{vir}/M_\odot = 1.5 \times 10^6 h^{-1} \sigma_v^3 $ in \citet{tul15} (adopting $h=0.7$), as listed in Table \ref{table2}: $M_{vir} = 2.1 \times 10^{14} M_\odot$ for the NGC\,4839 group, and  $M_{vir} = 3.0 \times 10^{14} M_\odot$ for the NGC\,4816 group.
We also list %the %virial 
%\textcolor{red}{the mass} 
the mass for the subhalo 2 corresponding to the NGC\,4816 group ($M_{vir} = 1.3 \times 10^{13} M_\odot$), and the subhalo 9  corresponding to the NGC\,4839 group ($M_{vir} = 1.7 \times 10^{13} M_\odot$) derived from the weak lensing analysis in \citet{oka14}. 
\citet{oka14} derived the mass within the truncation radius of each subhalo. 
The truncation radius of the subhalo 9 is 98 kpc, which is much smaller than the virial radius of the typical galaxy groups (the truncation radius of the subhalo 2 is not given in \citet{oka14}). 
Thus weak-lensing masses of the two groups are significantly smaller than the dynamical masses. 
%What causes this difference  is not clear at the moment, which may need further study. We note only the relative masses between the two groups derived using the same method.} 

These results show that the NGC\,4839 and NGC\,4816 groups have comparable high masses. 
This indicates that the NGC\,4816 group should show strong X-ray emission like the NGC\, 4839 group, but it is not yet detected in any previous X-ray observations.  Not all galaxy groups are detected in X-ray observations. About a half of the nearby galaxy groups show X-ray emission \citep{mul00}.  It is not clear why the NGC\, 4816 group does not show any strong X-ray emission, unlike the NGC\, 4839 group. It may need a study to investigate this issue further. %}

\subsection{CMDs of the GCs}

\begin{figure}[h]
    \centering
    \includegraphics[scale=0.37]{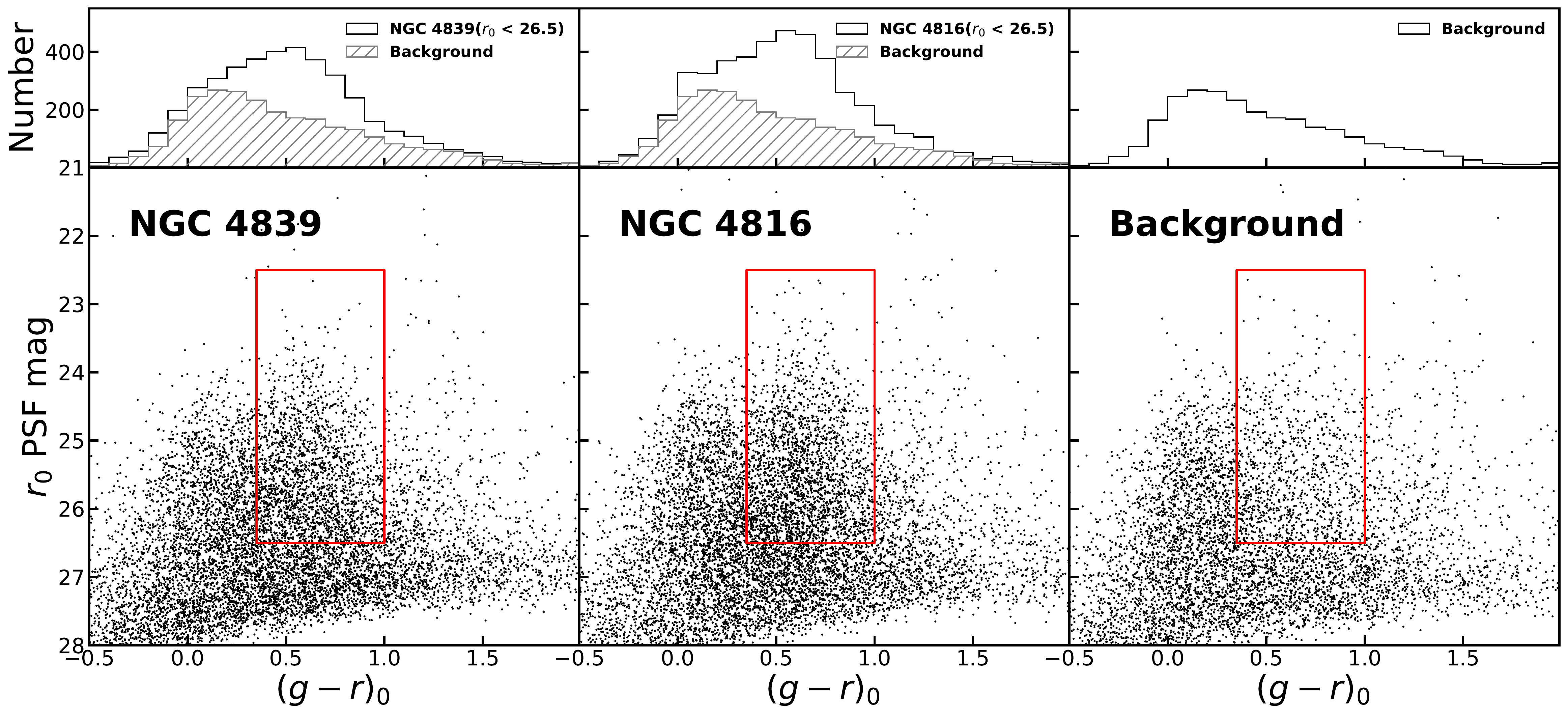}
    %{Figs1/CMD_N4839_N4816_back.pdf}
    \caption{Color-magnitude diagrams (lower panels) and color distributions (upper panels) of the point sources with $22.5<r_0<26.5$ mag 
    in the central regions  ($1'.2<R_{gal}<3'.3$) of NGC\,4839, NGC\,4816, and the background region with the same area based on the HSC images. 
    The hatched histograms in the upper panels for the galaxy regions represent the background region. 
    The red boxes in the lower panels represent the boundary for GC selection.
   }
    \label{fig:n4839_cmd} 
\end{figure}

In Figure \ref{fig:n4839_cmd} we plot the color-magnitude diagrams (CMDs) of the point sources  in the central regions ($R_{gal}<3\arcmin$ ($<87$~kpc)) of NGC\,4839  and  NGC\,4816 as well as a nearby background region with the same area as the galaxy region. 
We also plot the color histograms of the bright sources with $r_0<26.5$ mag of each region. 
To %derive 
show the net color histograms we %subtract 
display the contribution of the background sources in the galaxy regions using the background histogram.
%before and after background subtraction. 

The color histograms of the sources in the two galaxies  clearly show an excess (open histograms) with respect to the background region (hatched histograms) in the color range of $(0.0<(g-r)_0<1.3)$. 
In the CMDs the vertical structure seen inside the red box  represents mainly GCs in NGC\,4839 and NGC\,4816.
%, as also shown in the CMDs of other bright Coma galaxies in \citet{oh22}. 
We select GC candidates from the entire field using the color-magnitude criteria marked by the red box ($0.35<(g-r)_0<1.0$, and $22.5 <r_0<26.5$~mag) in the CMDs for the following analysis.

\subsection{Spatial Distribution of the GCs}

\begin{figure*}%[hbt!]
%    \centering
%    \includegraphics[scale=0.25]
%    \plottwo{Figs1/NGC4839_GC_sdm.pdf}
   % \includegraphics[scale=0.35]{fig2a.pdf} %Figs1/N4839_4816_findmap.pdf} 
    % \includegraphics[scale=0.6]{fig3.pdf} %fig2b.pdf} 
    %{Figs1/GC_field167_gal2.pdf}
    \includegraphics[scale=0.5]{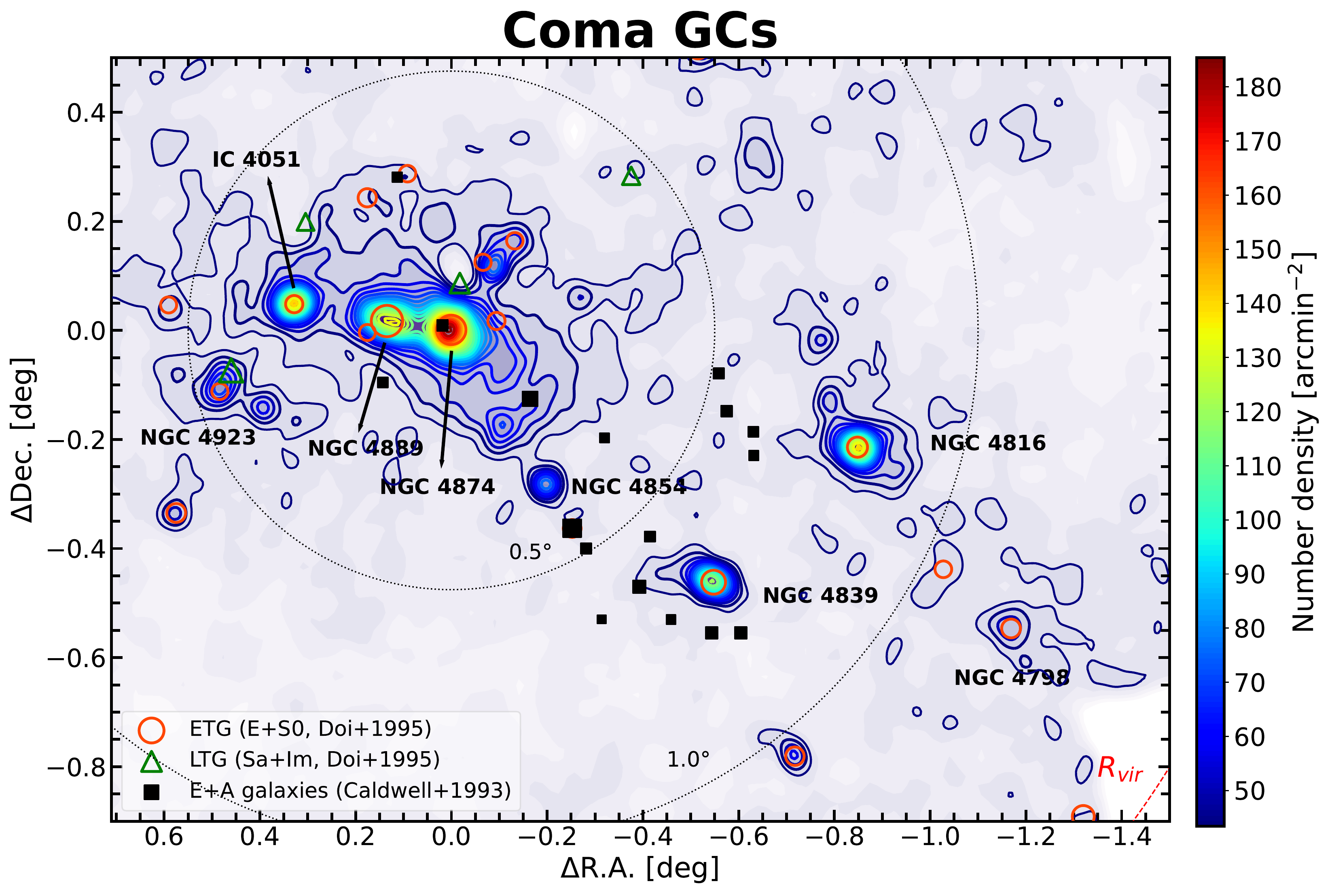} %rev2/GC_field167_gal2.pdf}
	\caption{
	%(a) A gray scale map of the $r$-band SDSS image of NGC\,4839 and its environment. Zoom-in fields (red boxes) are $10\arcmin \times 10\arcmin$.
	%(b) 
	Spatial number density contour map of GCs in the Coma field including NGC\,4839 and NGC\,4816 (see \citet{oh23} for details). 
	Dotted line  circles represent $R=0.5^\circ$, $1.0^\circ$, and ~$R_{vir}$(=2.8 Mpc) from NGC\,4874 at the  Coma center.
	%Hard to see the dotted line circles. Make them thicker!}
	Red circles and green triangles mark early-type galaxy members, and late-type galaxy members \citep{doi95}. Black boxes mark  E+A galaxies \citep{cal93}. The contour levels denote 2$\sigma_{bg}$ and %higher 
	larger with an interval of one $\sigma_{bg}$ where $\sigma_{bg}$ denotes the background fluctuation. The contour maps were smoothed using a Gaussian filter with $\sigma_G=1\arcmin$. The color bar represents the GC number density.
	}
	\label{fig:n4839_gcmap}
\end{figure*}

In Figure \ref{fig:n4839_gcmap} %1}(b) 
we display the spatial number density contour map of the selected GC candidates in NGC\,4839 and its environment.
The region covers out to the virial radius of Coma ($96\arcmin = 2.8$~Mpc), and NGC\,4839 and NGC\,4816 are located approximately at the half virial radius. 
The strongest peak of the GC number density is seen at the position of NGC\,4874, which is adopted as the Coma center in this study. Two other strong peaks in the main cluster core  are visible at the position of NGC\,4889 and IC 4051. % in the main cluster, and 
The main cluster core shows also a large extended distribution of intracluster GCs, the details of which are presented in \citet{oh23,lee22}. %Oh et al. (2022, Pape II).
Note that in the southwest outskirt two more strong peaks are found at the positions of NGC\,4839 and NGC\,4816, similar to those of NGC\,4889 and IC\,4051 in the main cluster core. 

One striking feature seen in this figure is a clear difference in the spatial distribution of GCs between NGC\,4839 and NGC\,4816: the spatial extent of the GC system in NGC\,4839 is very compact and that of the GC system in NGC\,4816 is much more extended, %although 
despite both galaxies showing a similarly strong peak at their centers.  
NGC\,4839 shows a weak excess tail of GCs in the east. There is no corresponding galaxies around the center of this excess. This may be due to stripped GCs from NGC\,4839. 
%\textcolor{red}{
\citet{sas16} noted that the center of the massive subhalo 9 in \citet{oka14} is  $1\arcmin$  east of NGC 4839 that is located  at the X-ray peak in the XMM-Newton and Suzaku images. This offset of the subhalo 9 might have also produced the east tail of the GCs in NGC 4839. %}

No bright galaxies are found in the outer region of NGC\,4816, which might have contributed  to the extended distribution of GCs. 
%\textcolor{red}{
The strong central concentration of the GCs around NGC\,4816 ($R<1200\arcsec$) in the radial number density profiles, as shown in the following section,  indicates that a majority of the GCs in this region are bound to the NGC\,4816 group. In the GC number density map of Figure~\ref{fig:n4839_gcmap} there are several weak GC clumps in the outskirts of the NGC\,4816 group, some of which can be due to some non-group member galaxies, but their contribution to the group GC system is negligible. %}

\subsection{Radial Number Density Profiles of the GCs}

 We derive the radial number density profiles of the GCs in NGC\,4839 and  NGC\,4816. We estimate the background levels from the surrounding regions (at $R_{gal}=9.2\arcmin$ %=8.35\arcmin-10.5\arcmin$ 
 for NGC\,4839 and $R_{gal}=26.4\arcmin$ for NGC\,4816), and subtract them from the original counts for the galaxy regions.  
 
\begin{figure*}%[hbt!]
    \centering
    %\plottwo{}{}
    %\includegraphics[scale=0.6]{fig4a.pdf} %NGC4839_gcRDP.pdf}
    \includegraphics[scale=0.6]{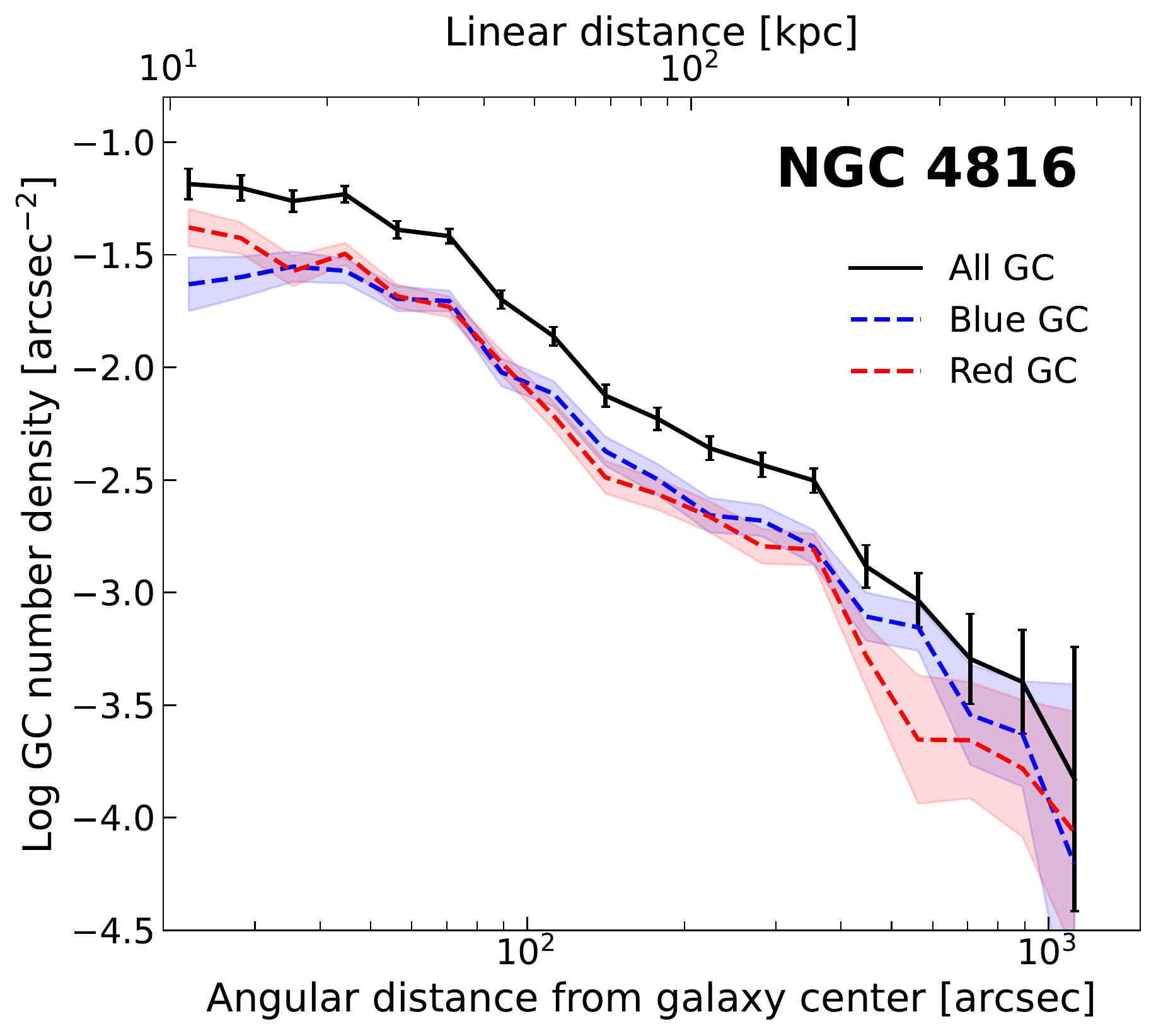}
    %rev2/NGC4839_gcRDP.pdf}
    %\includegraphics[scale=0.6]{fig4b.pdf} %NGC4816_gcRDP.pdf}
    \includegraphics[scale=0.6]{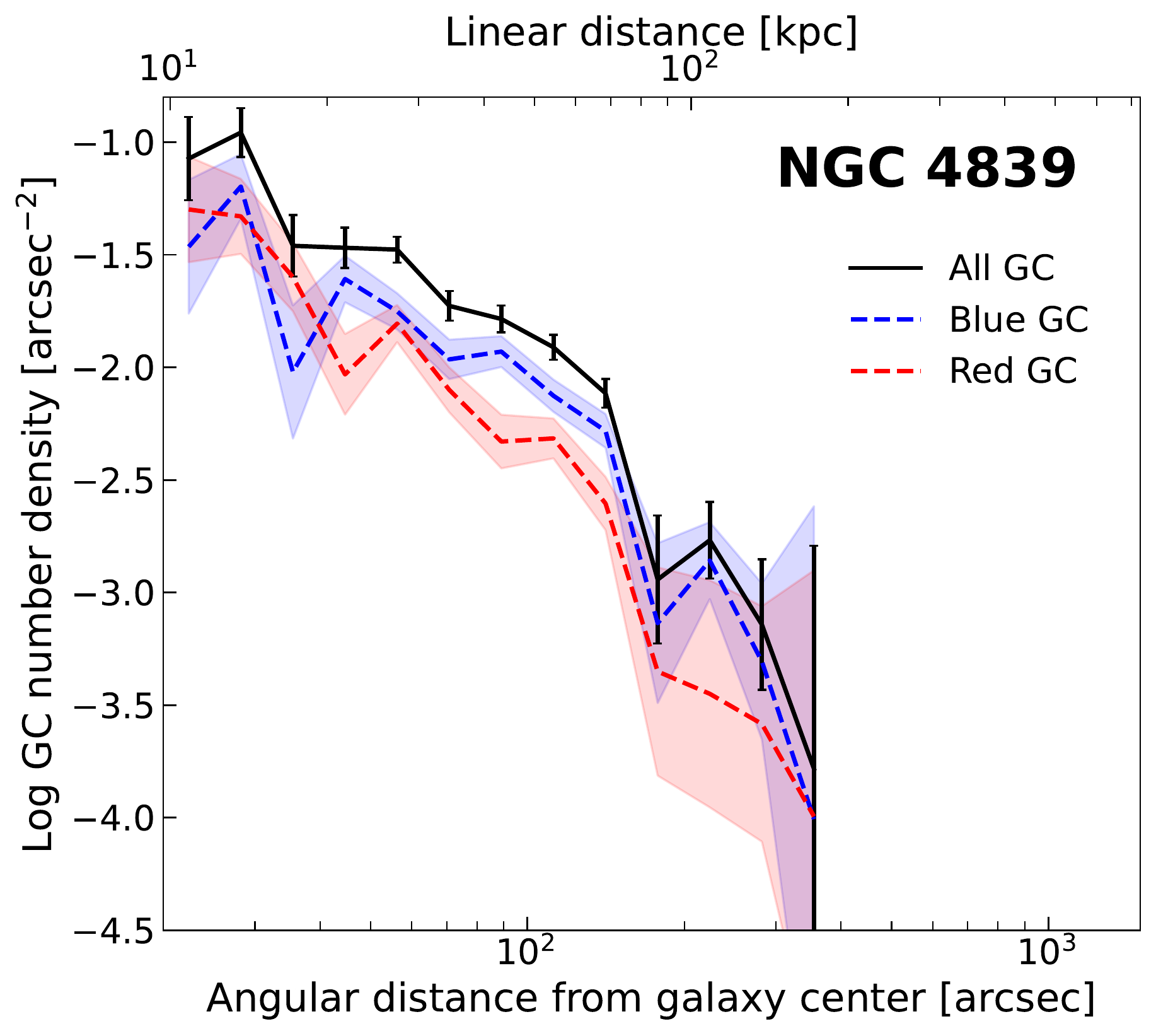} %rev2/NGC4816_gcRDP.pdf}
	\caption{
	Radial number density profiles of the GCs in NGC 4839 (upper panel) and NGC 4816 (lower panel): all GCs (black solid line), blue (metal-poor) GCs (blue dashed line), and red (metal-rich) GCs  (red dashed line).
	}
	\label{fig:n4839_n4816_rdp2}
\end{figure*}

 %\textcolor{red}{
 GC colors such as $(g-i)$ are a useful proxy for metallicity. The $(g-r)$ color in this study is less sensitive than the $(g-i)$ color, but is still useful. We divide the GC sample into two subsamples according to their color: blue (metal-poor) GCs with $0.35<(g-i)<0.655$, and the red (metal-rich) GCs with $0.655<(g-r)<1.0$, as described in \citet{oh23}. We derive the radial number density profiles of  the blue GCs and red GCs in NGC\,4839 and NGC
 ,4816, displaying them as well as that of all GCs in  Figure~\ref{fig:n4839_n4816_rdp2}. 
 This figure shows that the blue GC system is slightly more extended than the red GC system in both NGC\,4839 and NGC\,4816. %}

\begin{figure}[h]
    \centering
    %\plottwo{NGC_4839_SDSS.png}{NGC_4816_SDSS.png}
    %\plottwo{Figs1/NGC4839_galSBP_gcRDP.pdf}{Figs1/NGC4816_galSBP_gcRDP.pdf}
    %\includegraphics[scale=0.8]{revfigs/fig4.pdf}
    \includegraphics[scale=0.8]{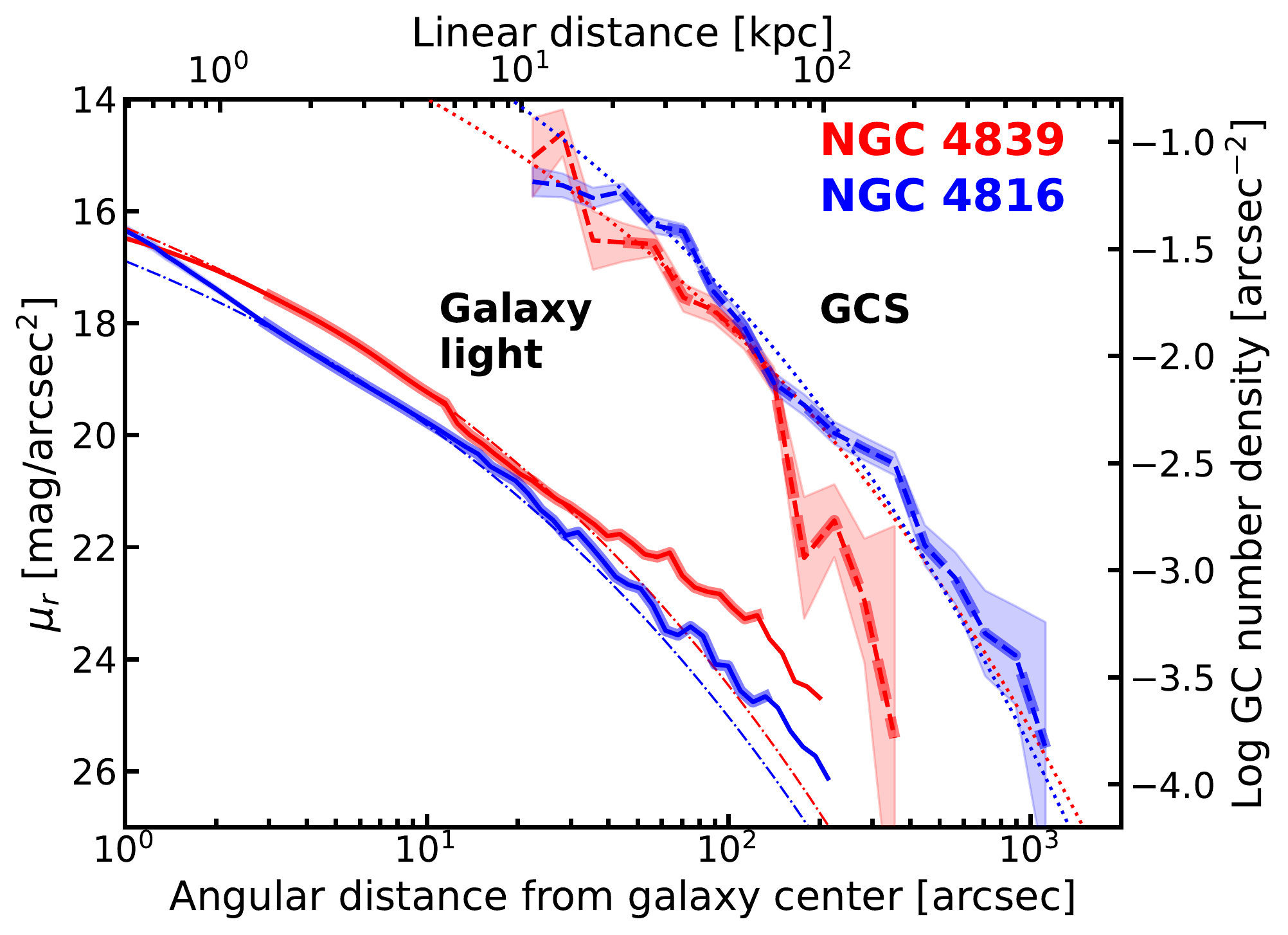} %NGC4839_4816_galSBP_gcRDP.pdf}
    \caption{
    Radial profiles for HSC $r$-band  surface brightness (solid lines) and GC number density (dashed lines) for NGC\,4839 (red lines) and NGC\,4816 (blue lines).
    Dot-dashed lines and dotted lines denote the results of S\'ersic law ($n=4$) fitting for galaxy light and GC number density profiles, respectively. Thicker lines denote the fitting ranges.
   }

    \label{fig:n4839_n4816_rdp}
\end{figure}

In Figure~\ref{fig:n4839_n4816_rdp}, 
%we plot the radial number density profiles of the GCs in NGC\,4839 in comparison with  NGC\,4816. 
we compare the radial profiles of the GC number density and surface brightness of galaxy light in NGC\,4839 and NG\,4816 
For comparison with galaxy light, we %also
derive the radial  surface brightness profiles of the two galaxies from the HSC $r$-band images.
First we mask out several bright sources except for the two galaxies in the images. Then we obtain surface brightness profiles of the galaxies using annular aperture photometry, and plot them in the same figure.

Several interesting features are noted in Figure \ref{fig:n4839_n4816_rdp}. %this figure.
First, the radial number density profiles of the GCs in the two galaxies show a striking difference in the outer region, while they are similar in the inner region. The decline in the central region at $R_{gal}<20\arcsec$ ($<10$~kpc) is due to incompleteness of our photometry, so we use only the data for the outer region at $R_{gal}>20\arcsec$. We note only the difference in the outer regions between the two galaxies. The radial number density profile of the NGC\,4839 GCs shows a sudden drop at $R_{N4839}\approx 80$~kpc, and few GCs are found at $R_{N4839} > 100$~kpc.
On the other hand, the radial number density profile of the NGC\,4816 GCs shows a slow decline even in the outer region at $R_{N4816} > 100$~kpc, and some GCs are found even out to $R_{N4816} \approx 500$~kpc.

Second, the surface brightness profiles of the two galaxies are similar in the inner region at $1<R_{gal}<20$~kpc, and show a slight difference in the outer region at $20<R_{gal}<50$ kpc.
The shapes of these profiles are also similar to that of the GC number density profile of NGC\,4816, but showing a clear difference against that of the GC number density profile of NGC\,4839.

Third, we fit the surface brightness profiles of the galaxies ($3\arcsec<R_{gal}<30\arcsec$) with a S\'ersic law for $n=4$ (i.e., a de Vaucouleurs law), as shown by the dot-dashed lines. The surface brightness profiles of the galaxy light in the inner regions of the two galaxies are reasonably fit by the S\'ersic law. 
The effective radius of the NGC\,4839 galaxy light,
$R_{\rm eff, N4839}=23\farcs5 \pm 0\farcs7 = 11.4\pm0.3$~kpc, is similar to that of the NGC\,4816 galaxy light, $R_{\rm eff, N4816}=23\farcs9 \pm 1\farcs1 = 11.6\pm0.5$~kpc. 
The surface brightness profile of NGC\,4839 shows a slight excess over the fitting line %S\'ersic component ($n=4$) 
at $R>1\arcmin$, which is a cD envelope, consistent with the previous results in \citet{sch88,ali14}. %(but at $R>2\arcmin$) based on SDSS DR7 data. 
On the other hand, this excess is much weaker in the case of NGC\,4816.
%, indicating that it may be closer to a D galaxy.

Fourth, we fit the radial number density profiles of GCs at $50\arcsec<R_{gal}<1260\arcsec$ in NGC\,4816 with a S\'ersic law for $n=4$, %(i.e., a de Vaucouleurs law), 
as shown by the dotted line in the figure. 
The radial number density profile of the NGC\,4816 GCs %in the entire range %$R_{gal}<200\arcsec$ 
is approximately fit by the S\'ersic law. 
The effective radius of the NGC\,4816 GC system derived from this fitting, is 
$R_{\rm eff,GCS}=124\pm37$~kpc. 
In the case of NGC\,4839, the radial number density profile of the  GCs in the inner region ($50\arcsec<R_{gal}<200\arcsec$) is roughly fit by the S\'ersic law, but the number density is significantly lower than the fitted line in the outer region at $R_{gal}>200\arcsec$.  
% The effective radius of the NGC\,4839 GC system derived from this fitting, is $R_{\rm eff,GCS}=479\arcsec\pm??$ = ?? kpc, but this value is clearly an overestimate. 
We derive the GC system effective radius of the two galaxies, %NGC\,4839 GC system 
from the cumulative radial distribution of GCs. We assume that the number density profile is flat in the central region ($R_{gal}<25\arcsec$) where our data is incomplete (see \citet{lee08} for the radial number density profile of M60 GCs)). 
%\textcolor{red}{
The effective radius of the GC system derived from this, is 
%$R_{\rm eff,GCS}=101\farcs54\pm3\farcs09$ = $49.08\pm1.49$~ kpc for NGC\,4839.
$R_{\rm eff,GCS}=101\farcs5\pm3\farcs1$ = $49.1\pm1.5$~ kpc for NGC\,4839.
We resample the radial density profiles from the data 1000 times, and repeat the same procedure to derive an effective radius from each profile. From this we obtain a standard deviation of resampled $R_{\rm eff,GCS}$ as a measuring error. Note that the true error must be larger than this error. 
Similarly we obtain 
%$R_{\rm eff,GCS}=331\farcs18\pm10\farcs81$ = $160.07\pm5.22$~kpc for NGC\,4816, 
$R_{\rm eff,GCS}=331\farcs2\pm10\farcs8$ = $160.1\pm5.2$~kpc for NGC\,4816, 
which is larger than, but consistent, within the error, with the value based on the fitting.
Thus the effective radius of the NGC\,4839 GC system
is about three times smaller than that of the NGC\,4816 GC system. %}

 \section{Discussion and Conclusion}
 
 Coma is an ideal target for investigating not only the general assembly process of galaxy clusters but also the details of the merging process including the infall phase of substructures.
 Various substructures related with the merging process in Coma were discovered in previous X-ray images (see \citet{bri92,whi93,neu01,san20,mir20,chu21} and references therein). Early studies based on X-ray observations suggested that the NGC\,4839 group is in the first phase of infall \citep{bri92,whi93}. Then \citet{bur94} presented a new scenario, based on hydro/N-body simulations, that {\it Coma already had a lunch} (the NGC\,4839 group) and the NGC\,4839 group is in the second infall, which can explain the optical, radio, and X-ray properties of Coma. Later \citet{col96} pointed out the shortcomings of the arguments in \citet{bur94}, and argued that NGC\,4839 is in the first phase of infall, based on dynamics of a large number of Coma galaxies. 
 Most of these substructures could be explained either in pre-merger scenarios or in post-merger scenarios, as summarized in Table~\ref{table1}.
 
Later \citet{lys19} noted two prominent features seen in the XMM-Newton and Chandra images of the NGC\,4839 group: a long (600~kpc) bent tail of cool gas of NGC\,4839, and a sheath of enhanced X-ray surface brightness due to hotter gas in the southwest, and tried SPH simulations to test both pre-merger and post-merger scenarios. They concluded that the post-merger scenario can explain better the observational results (X-ray brightness and temperatures) than the pre-merger scenario. According to this scenario (see their Fig.~8), the NGC\,4839 group began falling  to the main cluster from the northeast about 2~Gyr ago, passed the center about 1.6~Gyr ago, and  began the second infall after reaching the apocenter in the southwest recently.

 Recently from the X-ray images obtained with the SRG/eROSITA, \citet{chu21,chu22} found a faint X-ray bridge connecting the NGC\,4839 group with the main cluster. This bridge may be a remnant of stripped gas while NGC\,4839 moves outward from the main cluster to the current position, showing that it is %a 
 strong evidence that NGC\,4839 already passed the main cluster core  (see their Fig. 11).
 %If the NGC\,4839 group is its first infall phase, the presence of the bridge cannot be explained.
 \citet{chu21,chu22} also pointed out that the existence of the bow shock at $R\approx  33\arcmin$ (960~kpc) in the west and the radio relic at $R\approx 2.1$~Mpc in the southwest \citep{bon21} may correspond, respectively, to the secondary shock (produced when crossing the apocenter) and the primary shock (produced when crossing the main cluster core) caused by the merging event with NGC\,4839.

\begin{figure*}%[hbt!]
    \centering
    \includegraphics[scale=0.8]{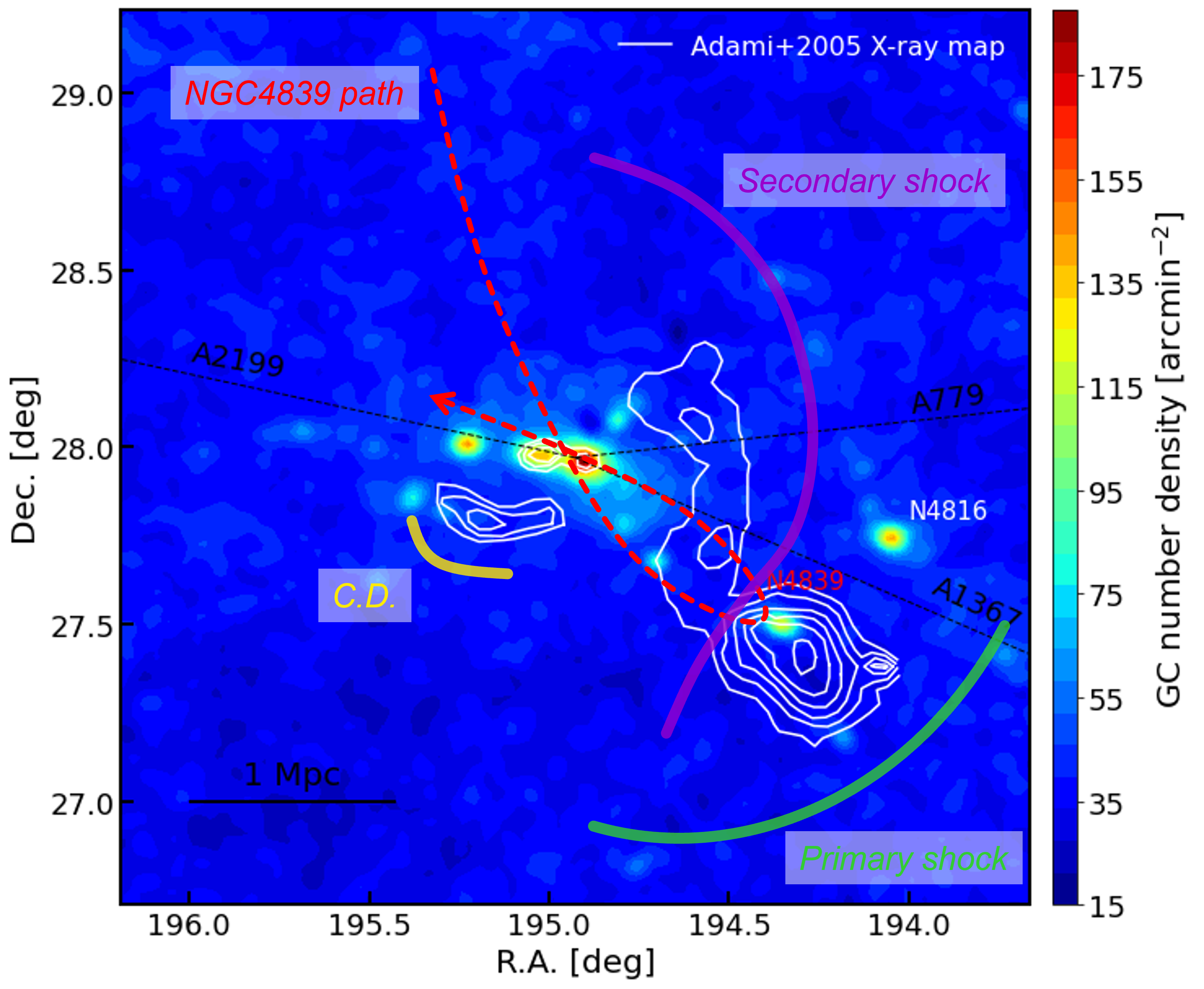} %Adami05_GCcontour_3_illust.png}
	\caption{Comparison of the GC number density map (pseudo color map) with 
	XMM-Newton X-ray 	map \citep{neu01,neu03} after $\beta$ model subtraction (white contours,
	%\citet{neu01}, 
	based on Figure 3 in \citet{ada05}). 
	Dotted black lines mark the direction of neighboring large scale structures.
	Green, purple, and yellow lines denote the primary shock, secondary shock, and contact discontinuity, respectively, in \citet{chu21} (from their Fig. 11). 
	The red dashed line shows the trajectory of the NGC\,4839 group suggested for the second-infall scenario \citep{lys19,chu21}. The color bar represents the GC number density.
	} 
	\label{fig:map1Adami05}
\end{figure*}

 In Figure~\ref{fig:map1Adami05}
 we compare  the GC number density map (pseudocolor map)  with
the XMM-Newton X-ray contour map of hot gas obtained after $\beta$ model subtraction (showing substructures  better, \citet{neu01,neu03}) (based on Fig.~3 in \citet{ada05}). 
%\textcolor{red}{
In this figure, the X-ray contours around the NGC 4839 region show a slight offset from the center of the NGC 4839 GC clump. This offset is not seen in the recent X-ray data \citep{lys19,chu21}. This offset is due to the outdated X-ray data \citep{neu03} used in \citet{ada05}. %} 
The X-ray map shows three prominent substructures:
(a) the NGC\,4839 group where a strong concentration of GCs is seen only at the position of NGC\,4839,
(b) a large arc-like western substructure where few GCs are found, and 
(c) a smaller substructure associated with NGC\,4911/4921 in the southeast where only a small population of GCs are seen. 
Note that the X-ray emission substructure is seen in the NGC\,4839 group, %while none
but not in the NGC\,4816 group.

%\textcolor{red}{
The center of the NGC\,4839 group (G2 in \citet{ada05}) was close to NGC\,4839 in the old study by \citet{ada05}. However, the recent study based on a much larger sample of Coma members by \citet{hea21} shows that the center of the NGC\,4839 group (S11) is significantly offset to the southwest from NGC\,4839 (see their Fig.~11). On the other hand, the recent SRG/eROSITA X-ray data with higher spatial resolution \citep{chu21} (as well as XMM-Newton data) shows clearly an X-ray peak at the position of NGC\,4839 which is embedded in a much more diffuse X-ray emission. This diffuse component is significantly overlapped with the galaxy distribution of the S11 group (see Fig. 11 in \citet{hea21}). %}

 In the figure we also add the trajectory (red dashed line) of the NGC\,4839 group suggested for the second-infall scenario\citep{lys19,chu21} (from Figure 11 in \citet{chu21}), as well as other known substructures.
The very compact spatial extent of the GC system in NGC\,4839, much smaller than the GC system in NGC\,4816, can be explained if NGC\,4839 lost a significant number of GCs in the outskirt of NGC\,4839 when it passed the main cluster. 

%\textcolor{red}{
On the other hand, the more extended GC system
in NGC\,4816 indicates that it may be in the first phase of infall, as described below.
The radial velocity of the NGC 4839 group is 768 \kms~~ larger than that of the main cluster (6853 \kms). \citet{col96} suggested that the angle between the observer and the velocity vector of the NGC\,4839 group is about 74 deg so the merger is happening with $\Delta v=1700$ \kms~ almost in the projected sky plane. Which of the main cluster and NGC\,4839 is closer to us is not yet known. 
On the other hand, the relative velocity of the  NGC\,4816 group %(6898 \kms) 
with respect to the main cluster  is only +35 \kms~ and the NGC\,4816 group is located along the large scale filament connecting with Abell\,1367. Considering these we infer that the NGC\,4816 group is infalling to  the cluster center in the sky plane. In addition, the GC system of the NGC\,4816 shows an extended structure with a continuously declining radial number density profile. These results indicate that the NGC\,4816 is in its first infall. If it is in its second infall, its radial profile of the GC system would have shown a significant drop in the outer region like the one in the NGC\,4839 group. %}

 If NGC\,4839 is in the first phase of infall, it should %is supposed to 
 show a similar distribution to that of NGC\,4816, and it would be difficult to explain the observed difference between NGC\,4839 and NGC\,4816.
 When NGC\,4839 crosses the main cluster core again, it would lose more GCs, which will become part of the intracluster GCs.
 
 In conclusion, the spatial distribution of GCs in NGC\,4839 and its environment supports %that the post-merger scenario.
 the second infall scenario where the NGC\,4839 passed the Coma center about 1.6~Gyr ago, and began the second infall after reaching the apocenter in the southwest. %recently.
 %\textcolor{red}{
 Previous simulations on GCs in galaxy clusters (e.g., \citet{ram18,ram20}) are useful to understand the spatial distribution and kinematics of the GCs in large scales. However none of them provide any results on how the motion of individual groups in galaxy clusters affects the size of the GC systems in individual galaxies, which could be compared with the results in this study. We expect that our results motivate future simulations to address this issue. %}

\acknowledgments

This work  was supported by the National Research Foundation grant funded by the Korean Government (NRF-2019R1A2C2084019). %Thanks to Jisu Kang.
We thank Brian S. Cho for his help in improving the English in the manuscript.
The authors are grateful to the anonymous referee for useful comments.
\vspace{5mm}
\facilities{Subaru(Hyper Suprime-Cam)}

%\software{SExtractor \citep{ber96}}



%\appendix

%\clearpage

%%%%%%%%%%%%%%%%%%%%%%%%%%%%%%
% Table 1  Merging-related features of the NGC\,4839
%%%%%%%%%%%%%%%%%%%%%%%%%%%%%%
\begin{deluxetable}{lllc}[]
	\setlength{\tabcolsep}{0.05in}
	%\rotate
	\tablecaption{Merging-related Features for the NGC\,4839 Group	\label{table:mfeatures}}
	\tablewidth{0pt}
	\tablehead{ 
		\colhead{Band} & 
		\colhead{Features} & 	\colhead{Infall Phase\textsuperscript{a}} &
		\colhead{Reference\textsuperscript{b}}}
	\startdata
	X-ray 	&  NGC\,4839 inner tail (SW), SW main tail (sheath)  	& 1,2 &  1,2	\\
		 	&  SW bridge connecting NGC\,4839 and the main cluster & 2 & \\
		 	&   W sharp edge, E contact discontinuity  	& 1,2 &	\\
	Radio (cont) 	& Coma radio halo, W halo front & 2 	& 3,4	\\
		 	&  SW bridge, SW streams, SW relic ($R=2.1$ Mpc)  & 2	& 	\\
		 	Radio (HI) 	&  HI deficiency and old galaxies in NGC\,4839 group  & 1,2 	& 5	\\
	Optical &  E+A galaxies in SW$^c$  	& 2   & 6 \\
	        & galaxy distribution and kinematics     & 1,2 & 7\\
	        & {\bf Compact globular cluster system in NGC\,4839} & 2 & This study
		\\
	\hline
	\enddata
	\tablenotetext{a}{1 for the first infall phase (a pre-merger scenario), and 2 for the second infall phase (a post-merger scenario).}
	\tablenotetext{b}{1.\citet{lys19};2.\citet{chu21,chu22}; 3. \citet{kim89}; 4.\citet{bon21,bon22};5.\citet{hea21};6.\citet{cal93}; 7.\citet{col96}. }
	\tablenotetext{c}{\citet{col96} pointed out that a few of the E+A galaxies that are the members of the NGC\,4839 group, and that these E+A galaxies may be falling recently into Coma like NGC\,4839.}
	\label{table1}
\end{deluxetable}

\clearpage
%%%%%%%%%%%%%%%%%%%%%%%%%%%%%%
% Table 2  The NGC\,4839 Group vs. the main cluster
%%%%%%%%%%%%%%%%%%%%%%%%%%%%%%
\begin{deluxetable}{lcccc}[]
	\setlength{\tabcolsep}{0.05in}
	%\rotate
	\tablecaption{%\textcolor{red}{
	Basic Parameters for the Main Cluster, NGC\,4839 and NGC\,4816 Group in Coma %}
	\label{table:info}}
	\tablewidth{0pt}
	\tablehead{ 
		\colhead{Parameter} & 
		\colhead{Main cluster} & 
		\colhead{NGC\,4839 group} & \colhead{NGC\,4816 group} & 
		\colhead{Reference$^a$}}
	\startdata
%		R.A.(J2000) 	& $12^h$ $59^m$ $48^s$ 	& & NED	\\
%	Decl.(J2000) 	& $27\arcdeg$ $58\arcmin$ $48\arcsec$   & & NED	\\
%	Virial Radius & 2.1 Mpc (1.2 deg) & \citet{chu21}\\
	Heliocentric galaxy velocity, $v_{h}$ & 7167 \kms  & 7338 \kms & 6915 \kms  & 1,2,3 \\
	Heliocentric group velocity, $v_{h}$ & 6853 \kms  & 7621 \kms & 6898 \kms & 1,2,3 \\
	Velocity dispersion, $\sigma_v$ & 1082 \kms & %329, 
	462 \kms  & 521 \kms & 2,3 \\
	%Virial Mass (dynamics)$^c$, $M_{vir}$ & $1.3\times 10^{15} M_\odot$ & $8.6\times 10^{12} M_\odot$ & & 2 \\
	Virial Mass (dynamics)$^b$, $M_{vir}$ & $2.7\times 10^{15} M_\odot$ & $2.1\times 10^{14} M_\odot$ &$3.0\times 10^{14} M_\odot$ & 4 \\
%	Virial Mass (WL)
 Weak Lensing Mass$^c$, $M_{WL}$ & $1.2\times 10^{15} M_\odot$ & $1.7\times 10^{13} M_\odot$ & $1.3\times 10^{13} M_\odot$ & 5\\
	\hline
	\enddata
	%\tablenotetext{a}{ \citet{col96} suggested that the angle between the observer and the velocity vector of the NGC\,4839 group  is about 74 deg so the merger is happening with $\Delta v=1700$ \kms~~ almost in the projected sky plane. Which of the main cluster and NGC\,4839 is closer to us is not yet known.	}
	\tablenotetext{a}{1: NED; 2: \citet{col96}; 3: \citet{hea21}; 4: This study; 5: \citet{oka14}. }
	%\tablenotetext{c}{Given for the velocity dispersion of the NGC\,4839 group, 329 \kms, in \citet{col96}.}
	\tablenotetext{b}{Calculated for the velocity dispersion  %(462 \kms~ for the NGC 4839 group) 
	\citep{hea21} using the group virial mass equation: $M_{vir}/M_\odot = 1.5 \times 10^6 h^{-1} \sigma_v^3 $ in \citet{tul15}. Note that \citet{col96} presented $M_{vir}=1.3\times 10^{15} M_\odot$ for the main cluster, and $M_{vir}=8.6\times 10^{12} M_\odot$ for the NGC 4839 group from galaxy dynamics.}
	\tablenotetext{c}{
 Projected 
 masses ($M_{2D}$) within the truncation radius 
 for the subhalo 2 for the NGC\,4816 group, and the subhalo 9  for the NGC\,4839 group derived from the weak lensing analysis in \citet{oka14}, given for $h=0.7$.}
	\label{table2}
\end{deluxetable}
%%%%%%%%%%%%%%%%%%%%%%%%%%%%%%
% Figure 1
%%%%%%%%%%%%%%%%%%%%%%%%%%%%%%

%%%%%%%%%%%%%%%%%%%%%%%%%%%%%%
% Figure 2
%%%%%%%%%%%%%%%%%%%%%%%%%%%%%%

%%%%%%%%%%%%%%%%%%%%%%%%%%%%%%
% Figure 3
%%%%%%%%%%%%%%%%%%%%%%%%%%%%%%

%%%%%%%%%%%%%%%%%%%%%%%%%%%%%%
% Figure 4
%%%%%%%%%%%%%%%%%%%%%%%%%%%%%%
\clearpage

%%%%%%%%%%%%%%%%%%%%%%%%%%%%%%
% Figure 5      4
%%%%%%%%%%%%%%%%%%%%%%%%%%%%%%
\clearpage

%%%%%%%%%%%%%%%%%%%%%%%%%%%%%%
% Figure 6       5
%%%%%%%%%%%%%%%%%%%%%%%%%%%%%%
\clearpage

%========Supplementary Material ==================
%== Supplementary figures ==

%\clearpage

\end{document}